\documentclass[conference]{IEEEtran}
\IEEEoverridecommandlockouts
\usepackage[utf8]{inputenc}

\usepackage{cite}
\usepackage{amsmath,amssymb,amsfonts}
\usepackage{algorithmic}
\usepackage{graphicx}
\usepackage{textcomp}
\usepackage{epstopdf}
\usepackage[font=small,skip=0pt]{caption}
\usepackage{multirow}
\usepackage[english]{babel}

\newtheorem{theorem}{Theorem}
\newtheorem{lemma}{Lemma}

\usepackage[usenames, dvipsnames]{color}
\newcommand{\snr}{\mathsf{SNR}}
\newcommand{\sep}{\mathsf{SEP}}

\newcommand{\defeq}{\ensuremath{\triangleq}} 
\newcommand{\dvo}{\mathsf{DVO}}
\newcommand{\re}[1]{#1_{\rm re}}
\newcommand{\im}[1]{#1_{\rm im}}

\newcommand{\Arg}[1]{\mathsf{Arg}\left(#1\right)}
\newcommand{\dist}[1]{\mathsf{dist}\left(#1\right)}
\newcommand{\paren}[1]{\left(#1\right)}
\newcommand{\sqparen}[1]{\left[#1\right]}
\newcommand{\brparen}[1]{\left\{#1\right\}}

\newcommand{\parenro}[1]{\left.\left[#1\right)\right.}
\newcommand{\abs}[1]{\left| #1\right|}
\newcommand{\field}[1]{\ensuremath{\mathbb{#1}}}
\newcommand{\R}{\ensuremath{\field{R}}} 
\newcommand{\C}{\ensuremath{\field{C}}} 
\newcommand{\ra}{\ensuremath{\rightarrow}} 
\newcommand{\PR}[1]{\ensuremath{\mathsf{Pr}\left\{#1\right\}}} 
\newcommand{\ES}[1]{\ensuremath{\mathsf{E}\left[#1 \right]}} 
\newcommand{\V}[1]{\ensuremath{\mathsf{Var}\left(#1 \right)}} 
\newcommand{\e}[1]{\ensuremath{{\rm e}^{#1}}} 

\def\BibTeX{{\rm B\kern-.05em{\sc i\kern-.025em b}\kern-.08em
    T\kern-.1667em\lower.7ex\hbox{E}\kern-.125emX}}
\begin{document}


\title{Phase Modulated Communication with Low-Resolution ADCs}

\author{

\IEEEauthorblockN{Samiru Gayan\IEEEauthorrefmark{1}, Hazer Inaltekin\IEEEauthorrefmark{2}, Rajitha Senanayake\IEEEauthorrefmark{1}  and Jamie Evans\IEEEauthorrefmark{1}\\}
\IEEEauthorblockA{\IEEEauthorrefmark{1}Department of Electrical and Electronic Engineering, University of Melbourne, Parkville, VIC 3010, Australia.\\}
\IEEEauthorblockA{\IEEEauthorrefmark{2}School of Engineering, Macquarie University, North Ryde, NSW 2109, Australia.\\}
\IEEEauthorblockA{E-mails:\IEEEauthorrefmark{1}hewas@student.unimelb.edu.au; \IEEEauthorrefmark{1}\{rajitha.senanayake, jse\}@unimelb.edu.au; \IEEEauthorrefmark{2}hazer.inaltekin@mq.edu.au}
}

\maketitle

\begin{abstract}
This paper considers a low-resolution wireless communication system in which transmitted signals are corrupted by fading and additive noise. First, a {\em universal} lower bound on the average symbol error probability ($\sep$), correct for all $M$-ary modulation schemes, is obtained when the number of quantization bits is not enough to resolve $M$ signal points. Second, in the special case of $M$-ary phase shift keying ($M$-PSK), the optimum maximum likelihood detector for equi-probable signal points is derived. Third, utilizing the structure of the derived optimum receiver, a general average $\sep$ expression for the $M$-PSK modulation with $n$-bit quantization is obtained when the wireless channel is subject to fading with a circularly-symmetric distribution. Finally, an extensive simulation study of the derived analytical results is presented for general Nakagami-$m$ fading channels. It is observed that a transceiver architecture with $n$-bit quantization is {\em asymptotically} optimum in terms of communication reliability if $n \geq \log_2M +1$. That is, the decay exponent for the average $\sep$ is the same and equal to $m$ with infinite-bit and $n$-bit quantizers for $n\geq \log_2M+1$. On the other hand, it is only equal to $\frac12$ and $0$ for $n = \log_2M$ and $n < \log_2M$, respectively. Hence, for fading environments with a large value of $m$, using an extra quantization bit improves communication reliability significantly.      
\end{abstract}


\section{Introduction}
\label{Section: Introduction}
Massive MIMO and millimeter wave (mmWave) communications are considered to be among the core technologies for next generation wireless networks since they can cope with the modern day demands of global mobile data traffic \cite{Andrews14, Hossain15, Akyildiz16}. In particular, they are well capable of providing high spectral efficiency targets required by emerging data intensive applications such as tele-health, autonomous driving and tactile Internet \cite{paper_40}.  However, with all the envisioned gains of wide-bandwidth multi-antenna communication systems, there still remains an important challenge of improving energy efficiency in next generation wireless networks. 

One main factor that increases energy consumption of a communication system is the use of high resolution analog-to-digital converters (ADCs) at transceivers \cite{paper_11}.  This is even more critical in massive MIMO systems, where the network elements (i.e., usually the base stations) are equipped with large numbers of radio frequency (RF) chains, and hence with many high-resolution ADCs. Furthermore, wider bandwidths require higher sampling rates to digitize analog signals due to sampling theorem \cite{book_5}. As the energy consumption by ADCs grows exponentially with their resolution level and linearly with their sampling rate \cite{paper_40, paper_12, paper_46}, using high speed, high resolution ADCs in a large antenna array will decrease the energy efficiency of a communication system exorbitantly. This renders practical implementations harder and alternative design approaches are desired. Using low-resolution ADCs, on the other hand, may provide a solution for this problem. However, to make this practical, it is first important to gain a comprehensive understanding about the optimum receiver structure with low-resolution quantizers and the resulting fundamental limits on communication performance. This is the goal of the current paper for phase modulated communication.  

We consider a simple but insightful single-antenna wireless communication system in which data transmission is corrupted by fading and noise. The analysis of the multi-antenna case is similar, where the receiver is augmented with an appropriate diversity combiner \cite{Brennan03}.  We first show the existence of an error floor below which the average symbol error probability ($\sep$) cannot be pushed down for any modulation scheme and quantizer structure. Then, focusing on phase modulated communication, we obtain the optimum maximum likelihood (ML) detector rule for equi-probable signal points. 

For phase modulation, a low-resolution ADC quantizes the phase of the received signal in such a way that only the information about the quantization region in which the received signal landed is sent to the detector.  Hence, this quantization process increases uncertainty about the transmitted signals, and is expected to result in higher $\sep$s. Surprisingly, our simulation results indicate the contrary {\em asymptotically} if enough number of bits is used for quantization. More formally, for $M$-ary phase shift keying ($M$-PSK) and Nakagami-$m$ fading, if the number of quantization bits $n$ is larger than or equal to $\log_2M + 1$, the decay exponent for the average $\sep$ is the same with the one achieved by an infinite-bit quantization, which is equal to $m$. On the other hand, it is equal to $\frac12$ for $n=\log_2M$ and $0$ for $n<\log_2M$. The observed ternary $\sep$ behavior is also verified by a general analytical expression derived for circularly-symmetric fading distributions.          

In \cite{paper_55}, Zhang {\em et al.} investigated the usage of low-resolution ADCs in communication systems from various angles such as detection, channel estimation and precoding. An ML detector for quantized distributed reception was presented in \cite{paper_3}, where the complexity of the detector grows exponentially with high signal constellations, number of transmit antennas and number of users in the uplink. To reduce implementation complexity, a near-ML detector was proposed in \cite{paper_8} by means of convex programming. Numerical examples show that the proposed near-ML detector is capable of performing well even with moderate size antenna arrays.

\textit{Notation}: We use uppercase letters to represent random variables and calligraphic letters to represent sets. We use $\R$ and $\R^2$ to denote the real line and $2$-dimensional Euclidean space, respectively. For a pair of integers $i \leq j$, we use $\sqparen{i:j}$ to denote the discrete interval $\brparen{i, i+1, \ldots, j}$. The set of complex numbers $\C$ is $\R^2$ equipped with the usual complex addition and complex multiplication. We write $z = \re{z} + \jmath \im{z}$ to represent a complex number $z \in \C$, where $\jmath = \sqrt{-1}$ is the {imaginary unit} of $\C$, and $\re{z}$ and $\im{z}$ denote {\em real} and {\em imaginary} parts of $z$, respectively.  
Every $z \in \C$ has also a {\em polar} representation $z = \abs{z}\e{\jmath \theta} = \abs{z}\paren{\cos\paren{\theta} + \jmath \sin\paren{\theta}}$, where $\abs{z} \defeq \sqrt{\re{z}^2 + \im{z}^2}$ is the magnitude of $z$ and $\theta = \Arg{z} \in [-\pi, \pi)$ is called the (principle) argument of $z$. As is common in the communications and signal processing literature, $\Arg{z}$ will also be called the phase of $z$ (modulo $2\pi$).  For a complex random variable $Z = \re{Z} + \jmath \im{Z}$, we define its mean and variance as $\ES{Z} \defeq \ES{\re{Z}} + \jmath \ES{\im{Z}}$ and $\V{Z} \defeq \ES{\abs{Z - \ES{Z}}^2}$, respectively.  We say that $Z$ is {\em circularly-symmetric} if $Z$ and $\e{\jmath \theta}Z$ induce the same probability distribution over $\C$ for all $\theta \in \R$ \cite{Picinbono94,Koivunen12}. 

\section{System Setup}
\label{Section: System Model}
\subsection{Channel Model and Signal Modulation} \label{Subsection: Channel Model}
We consider the classical point-to-point wireless channel model with flat-fading. For this channel, the received discrete-time baseband equivalent signal $Y$ can be expressed by   
\begin{equation}\label{eq1}
Y = \sqrt{\snr}H X + W,
\end{equation}
where $X \in \mathcal{C} \subset \C$ is the (normalized) transmitted signal, $\mathcal{C}$ is the signal constellation set, $\snr$ is the ratio of the transmitted signal energy to the additive white Gaussian noise (AWGN) spectral density, $H \in \C$ is the unit power channel gain between the transmitter and the receiver, and $W$ is the circularly-symmetric zero-mean unit-variance AWGN, i.e., $W \sim  \mathcal{CN}(0,1)$ \cite{book_5}. We note that the operational significance of $\snr$ in this model is its scaling of signal energy with respect to the noise power as a single system parameter. In order to formalize the receiver architecture and the optimum signal detection problem, we consider $\mathcal{C}=\brparen{\e{\jmath \pi \paren{\frac{2k +1}{M}-1}}}_{k=0}^{M-1}$ in the remainder of the paper, which is the $M$-PSK signal constellation.\footnote{This choice of $\mathcal{C}$ ensures that the phase of $X$ always lies in $[-\pi, \pi)$.} 

For ease of exposition, we only consider the case in which $M$ is an integer power of $2$. This is the common practical situation where the incoming information bits are first grouped together and then mapped to a signal point (for example by using Gray coding). Extensions of our results to the more general case of $M$ being any positive integer is straightforward, albeit with more complicated notation and separate analyses in some special cases.       

\subsection{Receiver Architecture} \label{Subsection: Receiver}
\begin{figure}[!t]
\begin{center}
\includegraphics[width=0.45\textwidth]{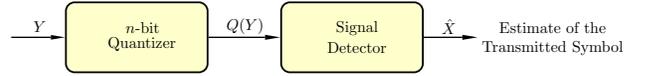}
\end{center}
\caption{An illustration of the receiver architecture with low-resolution quantization. The signal detector observes only the $n$-bit quantized versions of $Y$ to estimate the transmitted signal.}
\label{sys_model}
\end{figure}

The receiver architecture is based on a low-resolution ADC. As illustrated in Fig. \ref{sys_model}, this means that the received signal $Y$ first goes through a low-resolution quantizer, and then the resulting quantized signal information is used to determine the transmitted signal $X$. 
More specifically, if $n$ bits are used to quantize $Y$ before the detector, the quantizer $Q$ divides the complex domain $\C$ into $2^n$ quantization regions and outputs the index of the region in which $Y$ lies as an input to the detector. As such, we declare $Q(Y) = k$ if $Y \in \mathcal{R}_k$ for $k \in \sqparen{0:2^n -1}$, where $\mathcal{R}_k \subseteq \C$ is the $k$th quantization region.  Since information is encoded in the phase of $X$ with the above choice of constellation points, we choose $\mathcal{R}_k$ as the convex cone given by
\begin{eqnarray*}
\mathcal{R}_k = \brparen{z \in \C: \frac{2\pi}{2^n}k \leq \Arg{z} + \pi < \frac{2\pi}{2^n}\paren{k+1}},
\end{eqnarray*}
where $k \in \sqparen{0:2^n-1}$. 

We will assume that full channel state information is available at the receiver, and can be in turn used at the detector for signal recovery. This is achieved by using a high-resolution ADC (structured through either serial or parallel connections) during the channel estimation phase. The receiver then switches to low-resolution operation by using less number of quantization bits during the data transmission phase. Although increasing energy consumption, this is not a restrictive assumption in our case. Each fading state will span a large group of information bits at the target multiple Gbits per second data rates in $5$G and beyond wireless systems.  Hence, energy savings during data transmission are more significant than those during channel estimation.  

\section{Optimum Signal Detection} \label{Subsection: Signal Detection}
The aim of the detector is to minimize the $\sep$ by using the knowledge of $Q(Y)$ and channel state information, which can be represented as selecting a signal point $\hat{x}\paren{k, h}$ satisfying \begin{eqnarray*}
\hat{x}\paren{k, h} \in \underset{x \in \mathcal{C}}{\arg\max }\ \PR{X = x \big| Q(Y) = k, H = h}, 
\end{eqnarray*}
for  all $h \in \C$ and $k \in \sqparen {0: 2^n-1}$. The main performance figure of merit for the optimum detector is the average $\sep$ given by 
\begin{eqnarray}
p\paren{\snr} = \PR{X \neq \hat{x}\paren{Q\paren{Y}, H}}. \label{Eqn: SEP}
\end{eqnarray}

It is important to note that $p\paren{\snr}$, in addition to $\snr$, also depends on the number of quantization bits. Our first result indicates that there is an $\snr$-independent error floor such that the average $\sep$ values below it cannot be attained for $n<\log_2M$. The following theorem establishes this result formally. 

\begin{theorem} \label{Lemma: lower bound n<log2M}
Let $p_{\min}$ be the probability of the least probable signal point. If $n <\log_2 M$, then for any choice of modulation scheme and quantizer structure 
\begin{eqnarray}
p\paren{\snr}\geq  \frac{M-2^n}{2^n} p_{\min} \label{Eqn: Universal SEP Lower Bound}
\end{eqnarray} 
for all $\snr \geq 0$.  
\end{theorem}
\begin{IEEEproof}
See Appendix \ref{Appendix:Lemma: lower bound n<log2M}.
\end{IEEEproof}

We would like to highlight that the error floor in \eqref{Eqn: Universal SEP Lower Bound} is always a valid lower bound because $p_{\min} \leq \frac{1}{M}$. We also note that Fano's inequality can also be used to obtain similar, perhaps tighter, lower bounds on $p\paren{\snr}$ \cite{Cover91}. However, this will require the calculation of equivocation between $X$ and $Q(Y)$ for each modulation scheme and quantizer structure. Hence, it is not clear how to minimize over the modulation and quantizer selections in this approach. It is also important to note that the error floor in Theorem \ref{Lemma: lower bound n<log2M} is independent of the fading model. The unachievability of the average $\sep$ values below $\frac{M-2^n}{2^n} p_{\min}$ arises from the inherent inability of low-resolution ADC receivers to resolve different signal points when $n < \log_2M$. 

Next, we assume that all signal points in $\mathcal{C}$ are equiprobable, with probability $\frac{1}{M}$, and hence the optimum detector above is equivalent to the ML detector given by   
\begin{align}
\hat{x}\paren{k, h} &\in \underset{x \in \mathcal{C}}{\arg\max }\ \PR{Q(Y) = k \big| X = x, H = h} \label{Eqn: ML detector Q} 
\end{align}
for $h \in \C$ and $k \in \sqparen{0: 2^n-1}$. Given the events $\brparen{X = x}$ and $\brparen{H=h}$, we can write the probability $\PR{Q(Y) = k \big| X = x, H = h}$ as 
\begin{align}
    \PR{Q(Y) = k \big| X = x, H = h} = \nonumber \\
    &\hspace{-2.5cm} \int_{\mathcal{R}_k} \frac{1}{\pi} \exp\paren{-\abs{y - \sqrt{\snr}hx}^2} dy \label{Eqn: ML Probability}
\end{align} \label{Eqn: Prob of landing} 
\hspace{-0.4cm} since  $Y$ is conditonally a proper complex Gaussian random variable with mean $\ES{Y} = \sqrt{\snr}hx$ and variance $\V{Y} = 1$.  The integral in \eqref{Eqn: ML Probability} is with respect to the standard Borel measure on $\C$ \cite{Massey93}.  
We use the next lemma to describe a key result that is useful to establish the operation of the ML detector as given in the Theorem \ref{Theorem: ML Detector}. 

\begin{lemma} \label{Lemma: monotonically dec}
Let $\mathcal{R}$ be a convex cone given by $\mathcal{R} = \brparen{z \in \C: \alpha_1 \leq \Arg{z} \leq \alpha_2}$ for $\alpha_1,\alpha_2 \in \parenro{-\pi,\pi}$, and $W_1 \sim \mathcal{CN}\paren{\mu_1,1}$ and $W_2 \sim \mathcal{CN}\paren{\mu_2,1}$ be proper complex Gaussian random variables with means satisfying $\abs{\mu_1} = \abs{\mu_2}=r$ for some $r > 0$. Then, $\PR{W_1 \in  \mathcal{R}} \geq \PR{W_2 \in  \mathcal{R}}$ if $\abs{\mu_1 - z_{\rm mid}} \leq \abs{\mu_2 - z_{\rm mid}}$, where $z_{\rm mid} = r \e{\jmath \frac{\alpha_1 + \alpha_2}{2}}$.
\end{lemma}
\begin{IEEEproof}
See Appendix \ref{Appendix:Lemma: monotonically dec}   
 \end{IEEEproof}

We can use Lemma \ref{Lemma: monotonically dec} directly to obtain an ML detector rule. However, we find the next theorem more insightful to obtain an average $\sep$ expression for general fading distributions in the next section since it establishes the decision boundaries in terms of the bisectors of quantization regions.  

\begin{theorem} \label{Theorem: ML Detector}
Assume $H$ has a continuous probability density function. Then, $\hat{x}\paren{k, h}$ is unique with probability one, i.e., the set of $h$ values for which $\underset{x \in \mathcal{C}}{\arg\max }\ \PR{Q(Y) = k \big| X = x, H = h}$ is singleton has probability one, and the ML detection rule for the low-resolution ADC based receiver can be given as   
\begin{eqnarray}
\hat{x}\paren{k, h} = \underset{x \in \mathcal{C}}{\arg\min} \ \dist{\sqrt{\snr}hx, \mathcal{H}_k},
\end{eqnarray}
where $h \in \C$, $k \in \sqparen{0: 2^n-1}$, $\dist{z, \mathcal{A}}$ is the distance between a point $z \in \C$ and a set $\mathcal{A} \subseteq \C$, defined to be $\dist{z, \mathcal{A}} \defeq \inf_{s \in \mathcal{A}} \abs{z - s}$, and $\mathcal{H}_k = \brparen{z \in \C: \Arg{z} + \pi = \frac{\pi}{2^n}\paren{2k+1}}$.   
\end{theorem}
\begin{IEEEproof}
\begin{figure}[!t]
\center
\includegraphics[width=0.45\textwidth]{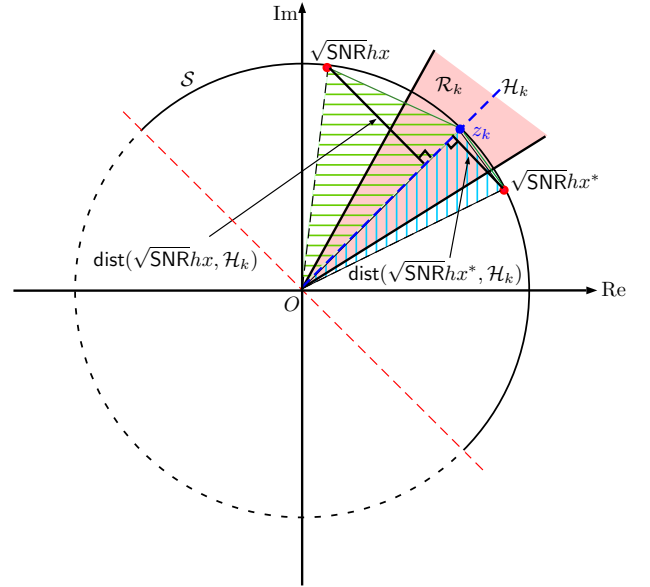}
\caption{An illustration for the proof of Theorem \ref{Theorem: ML Detector}, where $\abs{\sqrt{\snr}hx^\star - z_k} \leq \abs{\sqrt{\snr}hx - z_k}$. }
\label{Lemma1_Fig2}
\end{figure}
For $Q\paren{Y} = k$, the ML detector given in \eqref{Eqn: ML detector Q} reduces to finding a signal point in $\mathcal{C}$ maximizing the probability $\PR{\sqrt{\snr}hx + W \in \mathcal{R}_k}$, i.e., 
\begin{align}
    \hat{x}\paren{k, h} \in \underset{x \in \mathcal{C}}{\arg\max }\ \PR{\sqrt{\snr}hx + W \in \mathcal{R}_k}. \nonumber
\end{align}
By Lemma \ref{Lemma: monotonically dec}, $\hat{x}\paren{k,h}$ is the signal point in $\mathcal{C}$ such that $\sqrt{\snr}h \hat{x}\paren{k,h}$ is closest to $z_k = \sqrt{\snr}re^{\jmath \paren{\frac{2\pi}{2^n}k +\frac{\pi}{2^n}}}$, where $r = \abs{h}$. Further, $\hat{x}\paren{k,h}$ is unique with probability one due to the continuity assumption of the fading distribution. Consider now the semi-circle 
$$\mathcal{S} = \brparen{z \in \C : \abs{z} = \sqrt{\snr}r, \beta \leq \Arg{z} 
\leq \beta +\pi},$$
where $\beta = \paren{\frac{2\pi}{2^n}k + \frac{\pi}{2^n}-\frac{\pi}{2}}$. $\mathcal{S}$ is centered around $z_k$ and has $\mathcal{H}_k$ as its bisector, as illustrated in Fig. \ref{Lemma1_Fig2}. Let $x^\star \in \underset{x \in \mathcal{C}}{\arg\min} \ \dist{\sqrt{\snr}hx, \mathcal{H}_k}$.  For the $M$-PSK modulation scheme ($M\geq 2$) with regularly spaced signal points on the unit circle, we always have $\sqrt{\snr} h x^\star \in \mathcal{S}$ and $\sqrt{\snr} h \hat{x}(k,h) \in \mathcal{S}$. 

Take now another signal point $x \in \mathcal{C}$ different than $x^\star$ and satisfying $\sqrt{\snr} h x \in \mathcal{S}$. Consider the triangle formed by $O, z_k$ and $\sqrt{\snr} h x^\star$, and the one formed by $0, z_k$ and $\sqrt{\snr} h x$. We first observe that the area of the first triangle is smaller than the area of the second one since they share the line segment $\mathcal{L}_{Oz_k}$ as their common base but the height of the first one $\dist{\sqrt{\snr}hx^\star, \mathcal{H}_k}$ corresponding to this base is smaller than the height of the second one $\dist{\sqrt{\snr}hx, \mathcal{H}_k}$ corresponding to the same base. This is also illustrated in Fig. \ref{Lemma1_Fig2}. This observation, in turn, implies $\abs{\sqrt{\snr}hx^\star - z_k} \leq \abs{\sqrt{\snr}hx - z_k}$ because the remaining side lengths of both triangles are equal to $\sqrt{\snr} r$. Since this is correct for any $x \in \mathcal{C}$ satisfying $\sqrt{\snr} h x \in \mathcal{S}$, we conclude that $x^\star$ is unique and equal to $x^\star = \hat{x}(k, h)$.                 
\end{IEEEproof}

We note that the half-hyperplane $\mathcal{H}_k$ in Theorem \ref{Theorem: ML Detector} bisects the $k$th quantization region $\mathcal{R}_k$ into two symmetric regions. Hence, Theorem \ref{Theorem: ML Detector} indicates that the most probability mass is accumulated in the region $\mathcal{R}_k$ when the unit-variance proper complex Gaussian distribution with mean closest to $\mathcal{H}_k$ is integrated over $\mathcal{R}_k$, which coincides with the intuition.  

\section{Symbol Error Probability} 
\label{Section: SEP}
In this section, we will obtain a general $p\paren{\snr}$ expression for the optimum ML detector, which holds for any circularly-symmetric fading distribution.  We will only consider $n\geq \log_2 M$ due to the existence of an error floor for $n < \log_2 M$.
   
\begin{theorem}
\label{Theorem: p(snr) = 2^n P_00}
Let $H = R \e{\jmath \Lambda}$ be the circularly-symmetric fading coefficient with the joint phase and magnitude pdf $f_{R, \Lambda}\paren{r, \lambda} = \frac{1}{2\pi} f_R\paren{r}$ for $\lambda \in \parenro{-\pi, \pi}$ and $r \geq 0$. Then, $p\paren{\snr}$ is equal to
\begin{align}\label{Eqn: p(SNR) Eqn}
p\paren{\snr} &= \frac{2^{n-1}}{\pi} \int_{\frac{\pi}{M}-\frac{\pi}{2^n}}^{\frac{\pi}{M}+\frac{\pi}{2^n}}\int_{0}^{\infty}\PR{\sqrt{\snr}r \,\e{\jmath\theta} + W \notin \mathcal{E}}\nonumber \\
& \hspace{4.5cm} f_{R}\paren{r} \,dr \, d\theta, 
\end{align}
where $\mathcal{E} = \brparen{z \in \C: 0 \leq \Arg{z} < \frac{2\pi}{M}}$ and $\theta = \frac{\pi}{M} + \lambda$.
\end{theorem}
\begin{IEEEproof}
See Appendix \ref{Appendix:p(snr)=2^nP00 Proof}.
\end{IEEEproof}

An important remark about \eqref{Eqn: p(SNR) Eqn} is that $\sqrt{\snr}r \,\e{\jmath\theta} + W$ is a Rician distributed random variable. Hence, its joint and marginal distributions (phase and magnitude) are well studied in the literature \cite{Rice45, Cahn59, Dharmawansa18}. Specializing these results to our case, we can write its phase distribution $f_{\Phi}\paren{\phi}$ as in \eqref{Eqn: pdf of phase}, where $\mathcal{Q}\paren{\cdot}$ is the  complementary distribution function of the standard normal random variable. Then, we have   
\begin{figure*}
\begin{align}
    f_{\Phi}\paren{\phi}  &=  \frac{1}{2\pi}\e{-\snr r^2}\brparen{1 + 2\sqrt{\pi \snr}r\cos\paren{\phi - \theta}\e{\snr r^2 \cos^2\paren{\phi -\theta}}\mathcal{Q}\paren{-\sqrt{2\pi \snr}r\cos\paren{\phi - \theta}}} \label{Eqn: pdf of phase}
\end{align}
\hrule
\end{figure*}
\begin{align}
    \PR{\sqrt{\snr}r \,\e{\jmath\theta} + W \notin \mathcal{E}} = 1 - \int_0^{\frac{2\pi}{M}}f_\Phi \paren{\phi}\,d\phi,
\end{align}
which can be used to calculate $p\paren{\snr}$ in \eqref{Eqn: p(SNR) Eqn} numerically.  

\section{Numerical Results}
\label{Section: Numerical Results}
In this section, we present analytical and simulated average $\sep$ results for the $M$-PSK modulation with $n$-bit quantization.  Channel fading is circularly-symmetric with Nakagami-$m$ distributed magnitude. In order to characterize communication robustness, we focus on the decay exponent for $p\paren{\snr}$, which is given by\footnote{It can be shown that the limit in \eqref{Eqn: DVO Definition} always exists for this case.}
\begin{eqnarray}\label{Eqn: DVO Definition}
\dvo = - \lim_{\snr \ra \infty} \frac{\log{p\paren{\snr}}}{\log{\snr}}. 
\end{eqnarray}
Following the convention in the field, we will call $\dvo$ diversity order, although there is only a single diversity branch in our system. It should be noted that Nakagami-$m$ amplitude distribution can be obtained as the envelope distribution of $m$ independent Rayleigh faded signals for integer values of $m$ \cite{Nakagami60}. Hence, visualizing a Nakagami-$m$ wireless channel as a pre-detection analog square-law diversity combiner will help to put some of the observations below into context.           

\begin{figure}[!t]
\center
\includegraphics[width=0.48\textwidth]{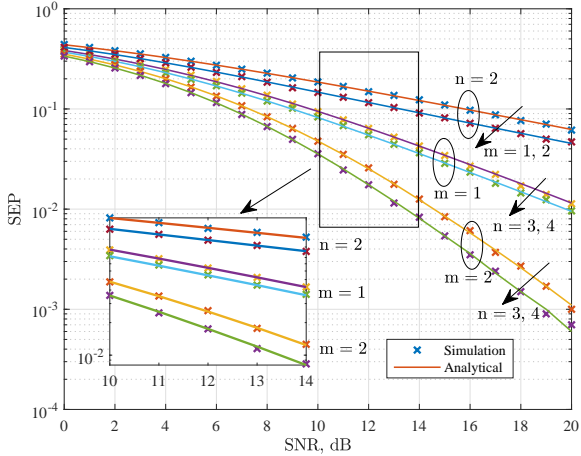}
\caption{Average SEP curves as a function of $\snr$ for QPSK modulation. $n=2,3,4$ and $m=1,2$.} 
\label{example1}
\end{figure}

Figure \ref{example1} plots the average $\sep$ as a function of $\snr$ for QPSK modulation with $n=2,3,4$-bit quantization under Nakagami-$m$ fading with shape parameter $m=1$ and $2$. The simulated results are generated by using Monte Carlo simulations, while the analytical results are obtained by using Theorem \ref{Theorem: p(snr) = 2^n P_00}. 
As the plot illustrates, the analytical results accurately follow the simulated results for all cases. We observe a noteworthy improvement in average $\sep$ when $n$ changes from $2$ to $3$ bits for QPSK modulation for both $m=1$ and $2$. Indeed, the jump in the average $\sep$ performance with one extra bit, on top of $2$ bits, is an improvement in $\dvo$ from $\frac12$ to $m$.  This can be seen through simple linear curve fitting. We also observe that the average $\sep$ reduces as we increase $n$, but the amount by which it reduces also gets smaller with increasing $n$. This can be clearly observed from the zoomed-in section in Fig. \ref{example1}. There is no change in $\dvo$ after $n \geq 3$, which is equal to $m$. There is also no change in $\dvo$ with $m$ for $n=2$, which is equal to $\frac12$.

\begin{figure}[!t]
\center
\includegraphics[width=0.48\textwidth]{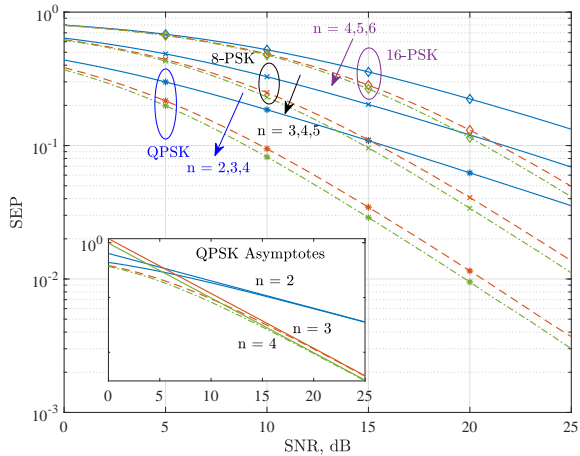}
\caption{Average SEP curves as a function of $\snr$ for different modulation schemes. $n=\log_2M, \log_2M + 1, \log_2M + 2$ and $m=1$.} 
\label{example2}
\end{figure}

Figure \ref{example2} plots the average $\sep$ as a function of $\snr$ for QPSK, $8$-PSK and $16$-PSK modulations while keeping the Nakagami-$m$ shape parameter fixed at $m=1$, which is the classical Rayleigh fading scenario. We plot the average $\sep$ for each modulation scheme by using $n=\log_2M$, $\log_2M+1$ and $\log_2M+2$ bits. From the plots, we can clearly observe that QPSK with $2$-bit, $8$-PSK with $3$-bit and $16$-PSK with $4$-bit quantizations have a $\dvo$ of $\frac{1}{2}$. Further, we can observe that QPSK with $3$ or more bits, $8$-PSK with $4$ or more bits, $16$-PSK with 5 or more bits quantizations have a $\dvo$ of $1$, which is equal to $m$ in this case. To further emphasize this point, the zoomed-in section in Fig. \ref{example2} illustrates the asymptotic average $\sep$ versus $\snr$ for QPSK modulation. These numerical observations indicate a ternary behavior in the decay exponent for $p\paren{\snr}$ depending on whether $n \geq \log_2M+1$, or $n = \log_2M$, or $n < \log_2M$. The complete analytical justification of this result is involved, and hence omitted due to space limitations.

\begin{figure}[!t]
\center
\includegraphics[width=0.48\textwidth]{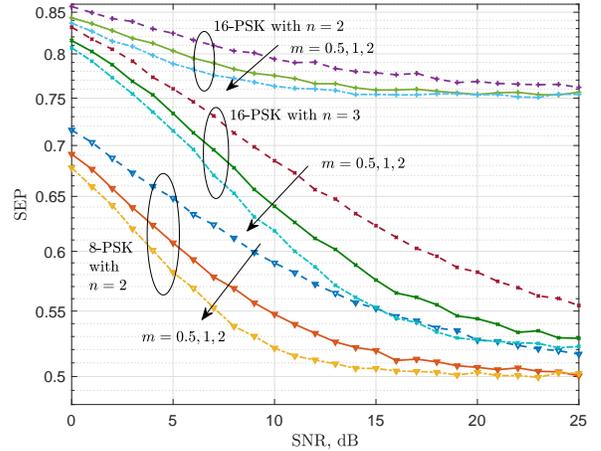}
\caption{Average $\sep$ as a function of $\snr$ for $8$-PSK and $16$-PSK modulations. $n=2<\log_2M$ and $m=1$.} 
\label{SEP_less_bits}
\end{figure}

Finally, to illustrate the error floor behavior obtained in Theorem \ref{Lemma: lower bound n<log2M}, we plot the simulated average $\sep$ curves as a function of $\snr$ for $8$-PSK and $16$-PSK modulations with $2$-bit quantization in Fig. \ref{SEP_less_bits}. All signal points are equi-probable. The channel model is the Rayleigh faded wireless channel, obtained by setting $m=1$.  The simulated results are again generated by using Monte Carlo simulations. We can clearly observe an error floor for high $\snr$ values when $n < \log_2M$ in Fig. \ref{SEP_less_bits}, as established by Theorem \ref{Lemma: lower bound n<log2M}. In particular, the average $\sep$ for $8$-PSK has a lower bound of $0.5$ with $2$-bit quantization. Similarly, the average $\sep$ for $16$-PSK has a lower bound of $0.75$ with $2$-bit quantization and a lower bound of $0.5$ with $3$-bit quantization. It should be noted that the error floor given in Theorem \ref{Lemma: lower bound n<log2M} is more conservative than those observed in Fig. \ref{SEP_less_bits}. This is because it is a universal lower bound that holds for all modulation schemes, quantizer types and fading environments, not only for very specific ones used to plot average $\sep$ curves in Fig. \ref{SEP_less_bits}.      

\section{Conclusions}
\label{Section: Conclusion}
We have obtained fundamental performance limits, optimum ML detectors and associated average $\sep$ expressions for low-resolution ADC based communication systems. We have also performed an extensive numerical study to illustrate the accuracy of the derived analytical expressions. A ternary $\sep$ behavior has been observed, indicating the sufficiency of $\log_2M+1$ bits for achieving  asymptotically optimum $M$-ary communication reliability. In most parts of the paper, we have focused on phase modulated communications. 

Phase modulation has an important and practical layering feature enabling the quantizer and detector design separation in low-resolution ADC communications.  For a given number of bits, the quantizer needs to be designed only once, and can be kept constant for all channel realizations. The detector can be implemented digitally as a table look-up procedure using channel knowledge and quantizer output. On the other hand, this feature is lost in joint phase and amplitude modulation schemes such as QAM. The quantizer needs to be dynamically updated for each channel realization in low-resolution ADC based QAM systems. This is because the fading channel amplitude may vary over a wide range, but the phase always varies over $\parenro{-\pi, \pi}$. However, phase modulation is historically known to be optimum only up to modulation order $16$ under peak power limitations \cite{Lucky62}. Hence, it is a notable future research direction to extend the results of this paper to higher order phase and amplitude modulations by taking practical design considerations into account. Similarly, utilizing the results of this paper, a detailed study on the receiver architecture design to determine where to place the diversity combiner (before or after quantizer or detector) and its type is needed when multiple diversity branches are available for data reception.                  

\appendices

\section{Proof of Theorem \ref{Lemma: lower bound n<log2M}}
\label{Appendix:Lemma: lower bound n<log2M}
In this appendix, we present the proof of Theorem \ref{Lemma: lower bound n<log2M}. We consider a class of hypothetical genie-aided detectors equipped with the extra knowledge of channel noise $W$. To this end, we let $g:\C^2 \times \sqparen{0:2^{n-1}}  \to \sqparen{0:M-1}$ be a genie-aided detector that has the knowledge of channel noise $W \in \C$, fading coefficient $H \in \C$ and quantizer output $Q\paren{Y} \in \sqparen{0:2^{n-1}}$. We also let $\mathcal{S}_{w, h, k} = \brparen{x \in \mathcal{C}: \sqrt{\snr}hx + w \in \mathcal{R}_k}$ be the set of signal points resulting in $Q(Y) = k$ for particular realizations of $H=h$ and $W=w$. We first observe that since $n<\log_2M$, there exists at least one quantization region $\mathcal{R}_{\tilde{k}}$ (depending on $w$ and $h$) such that $\mathcal{S}_{w, h, \tilde{k}}$ contains at least $\frac{M}{2^n}$ signal points. We note that $\frac{M}{2^n}$ is always an integer greater than $2$ since $M$ is assumed to be an integer power of $2$. Then, the conditional $\sep$ of any detector $g$ given $\brparen{W=w}$ and $\brparen{H=h}$, which we will denote by $p_g\paren{\snr,h,w}$, can be lower-bounded as 
\begin{align}
p_g\paren{\snr,h,w} &\nonumber \\
&\hspace{-2.1cm}\geq p_{\min} \hspace{-.3cm}\sum_{x \in \mathcal{S}_{w,h,\tilde{k}}}\hspace{-.3cm}\PR{g\paren{h,w,\tilde{k}}\neq x \, \big| \, W=w, H=h, X=x} \nonumber \\
&\hspace{-2.1cm}\geq \frac{M-2^n}{2^n} p_{\min}. \label{Eqn: Universal SEP Lower Bound Appendix}
\end{align}

By averaging with respect to $w$ and $h$, we also have $p_g\paren{\snr} \geq \frac{M-2^n}{2^n}p_{\min}$, where $p_g\paren{\snr}$ is the average $\sep$ corresponding to detector $g$. This concludes the proof since the obtained lower bound does not depend on the choice of modulation scheme, quantizer structure and detector rule. 

\section{Proof of Lemma \ref{Lemma: monotonically dec}}
\label{Appendix:Lemma: monotonically dec}
It is enough to show this result only for $\alpha_2 = -\alpha_1 = \alpha$. Otherwise, we can first rotate $W_1$, $W_2$ and $\mathcal{R}$ with $\e{-\jmath \frac{\alpha_1 + \alpha_2}{2}}$ and repeat the same analysis. Let $g\paren{\mu_i} = \PR{W_i \in \mathcal{R}}$ for $i=1,2$, and assume $\abs{\mu_1 - z_{\rm mid}} \leq \abs{\mu_2 - z_{\rm mid}}$. There are multiple cases in which the inequality $\abs{\mu_1 - z_{\rm mid}} \leq \abs{\mu_2 - z_{\rm mid}}$ holds, depending on $\mu_1$ and $\mu_2$ being located in the inside or outside of $\mathcal{R}$. We will consider only one case below due to space limitations. The analysis for other cases is similar.        
 
\begin{figure}[!t]
\center
\includegraphics[width=0.48\textwidth]{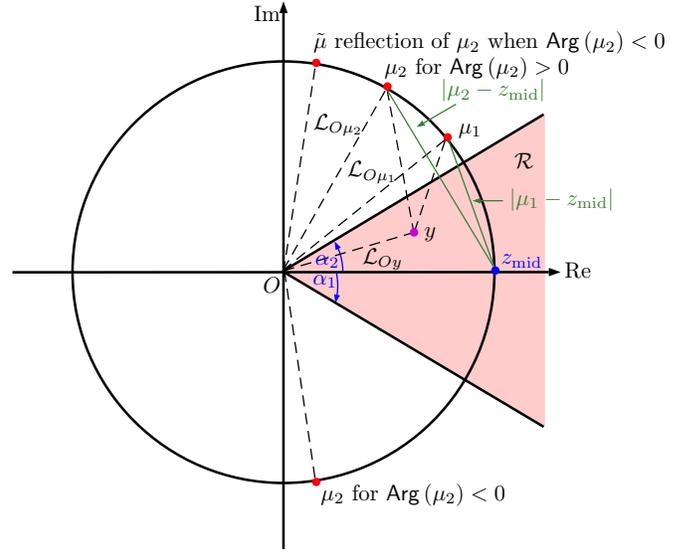}
\caption{An illustration for the proof of Lemma \ref{Lemma: monotonically dec} when $\mu_1$ and $\mu_2$ lie outside $\mathcal{R}^\circ$. $\abs{\mu_1}=\abs{\mu_2} = r$, $\abs{\mu_1 - z_{\rm mid}} \leq \abs{\mu_2 - z_{\rm mid}}$ and $\alpha_2 = -\alpha_1 = \alpha$.}
\label{Lemma1_1}
\end{figure}
 
To this end, we assume that both $\mu_1$ and $\mu_2$ lie outside $\mathcal{R}^\circ$, where $\mathcal{R}^\circ$ is the set of interior points of $\mathcal{R}$. This is the case shown in Fig. \ref{Lemma1_1}.  To start with, we will assume $0 \leq \Arg{\mu_1} \leq \Arg{\mu_2} < \pi$. Then, for any $y \in \mathcal{R}$, the angle between the line segments $\mathcal{L}_{Oy}$ and $\mathcal{L}_{O\mu_1}$ is smaller than the one between the line segments $\mathcal{L}_{Oy}$ and $\mathcal{L}_{O\mu_2}$.\footnote{The line segment $\mathcal{L}_{z_1z_2}$ between the points $z_1 \in \C$ and $z_2 \in \C$ is the set given by $\mathcal{L}_{z_1z_2} = \brparen{(1-t) z_1 + t z_2: t \in \sqparen{0,1}}$.} This is illustrated in Fig. \ref{Lemma1_1}, too. Hence, applying the cosine rule for the triangle formed by $O, y$ and $\mu_1$, and for the triangle formed by $O, y$ and $\mu_2$, it can be seen that $\abs{y - \mu_1} \leq \abs{y - \mu_2}$ for all $y \in \mathcal{R}$.\footnote{This statement is correct even when both $y$ and $\mu_1$ lies on the boundary of $\mathcal{R}$ and the triangle formed by $O, y$ and $\mu_1$ reduces to a line segment.} Therefore, $g\paren{\mu_1} = \frac{1}{\pi}\int_{\mathcal{R}} \exp \paren{-\abs{y-\mu_1}^2 } dy \geq \frac{1}{\pi}\int_{\mathcal{R}} \exp \paren{-\abs{y-\mu_2}^2 } dy = g\paren{\mu_2}$. Next, we assume $\Arg{\mu_2} \in \parenro{-\pi, 0}$ and $0 \leq \Arg{\mu_1} \leq \abs{\Arg{\mu_2}} \leq \pi$. Let $\widetilde{W}$ be the auxiliary random variable distributed according to $\mathcal{CN}\paren{\tilde{\mu} ,1}$ with $\tilde{\mu} = r\e{\jmath \abs{\Arg{\mu_2}}}$, i.e., $\tilde{\mu}$ is the reflection of $\mu_2$ around the real line. Symmetry around the real line implies that $g\paren{\mu_2}$ is equal to $ g\paren{\tilde{\mu}}= \PR{\widetilde{W} \in \mathcal{R}}$, which is less than $g\paren{\mu_1}$ due to our arguments above. For $\Arg{\mu_1} \in \parenro{-\pi, 0}$, the same analysis still holds after reflecting $\mu_1$ around the real line, leading to $g\paren{\mu_1} \geq g\paren{\mu_2}$ for all $\mu_1, \mu_2 \notin \mathcal{R}^\circ$ satisfying $\abs{\mu_1 - z_{\rm mid}} \leq \abs{\mu_2 - z_{\rm mid}}$.  
 
\section{Proof of Theorem \ref{Theorem: p(snr) = 2^n P_00}} 
\label{Appendix:p(snr)=2^nP00 Proof}
We consider a partition $\brparen{\mathcal{D}_k}_{k=0}^{2^n-1}$ of $\C$, where each element of this partition is given by $\mathcal{D}_k = \brparen{z \in \C: \paren{2k -1}\frac{\pi}{2^n} \leq \Arg{z} +\pi < \paren{2k+1}\frac{\pi}{2^n}}$ for $k \in \sqparen{1:2^n-1}$ and $\mathcal{D}_0 = \mathcal{D}_0^1 \bigcup \mathcal{D}_0^2$, where $\mathcal{D}_0^1 = \brparen{z \in \C: \pi - \frac{\pi}{2^n} \leq \Arg{z} < \pi}$ and $\mathcal{D}_0^2 = \brparen{z \in \C: -\pi \leq \Arg{z} < \frac{\pi}{2^n} - \pi}$. Let $x_i = \e{\jmath \pi\paren{\frac{2i+1}{M} - 1}}$ be the $i$th signal point for $i \in \sqparen{0:M-1}$. Then, we can express $p\paren{\snr}$ according to    
\begin{align}
p\paren{\snr} =\nonumber \\
& \hspace{-1cm}\frac{1}{M} \sum_{i=0}^{M-1}\sum_{k=0}^{2^n-1} \int\limits_{\mathcal{D}_k}\PR{x_i \neq \hat{x}\paren{Q\paren{Y},h}|H=h, X=x_i} \nonumber \\
& \hspace{4cm}f_H\paren{h} dh. \label{Eqn: p(SNR) 2}
\end{align}

We will show that all the terms in \eqref{Eqn: p(SNR) 2} are equal. To this end, we first define $\mathcal{E}_i = \brparen{z \in \C: \Arg{x_i} - \frac{\pi}{M} \leq \Arg{z} < \Arg{x_i} + \frac{\pi}{M}}$ for $i \in \sqparen{0: M-1}$. $\mathcal{E}_i$ contains all $\mathcal{H}_k$'s (i.e., bisectors of quantization regions) to which $x_i$ is the closest signal point. Furthermore, this statement continues to be true for $\sqrt{\snr}hx_i$ as long as $\Arg{h} \in \parenro{-\frac{\pi}{2^n}, \frac{\pi}{2^n}}$ since the angular spacing between $\mathcal{H}_k$'s is uniform and equal to $\frac{2\pi}{2^{n}}$. Notice that $\Arg{h} \in \parenro{-\frac{\pi}{2^n}, \frac{\pi}{2^n}}$ if and only if $h \in \mathcal{D}_{2^{n-1}}$. On the other hand, if $\Arg{h} \in \parenro{\frac{\pi}{2^n}, \frac{3 \pi}{2^n}}$, the region $\e{\jmath \frac{2\pi}{2^{n}}}\mathcal{E}_i$ contains all $\mathcal{H}_k$'s to which $\sqrt{\snr}hx_i$ is closest. Notice also that $\Arg{h} \in \parenro{\frac{\pi}{2^n}, \frac{3 \pi}{2^n}}$ if and only if $h \in \mathcal{D}_{2^{n-1}+1}$. Similarly, $\e{-\jmath \frac{2\pi}{2^{n}}}\mathcal{E}_i$ contains all $\mathcal{H}_k$'s to which $\sqrt{\snr}hx_i$ is closest if $\Arg{h} \in \parenro{-\frac{3\pi}{2^n}, -\frac{\pi}{2^n}}$, and $\Arg{h} \in \parenro{-\frac{3\pi}{2^n}, -\frac{\pi}{2^n}}$ if and only if $h \in \mathcal{D}_{2^{n-1}-1}$. The same idea extends to any $\mathcal{D}_k$, and we define
\begin{eqnarray}
\mathcal{E}_{i,k} \defeq \exp\paren{\jmath\paren{k-2^{n-1}}\frac{2\pi}{2^n}} \mathcal{E}_i, \label{Eqn: Equal Probability Sets}
\end{eqnarray}
for $i \in \sqparen{0:M-1}$ and $k \in \sqparen{0:2^n-1}$.     

To complete the proof, we let $p_{i,k}$ be the integral term in \eqref{Eqn: p(SNR) 2} 
for $i \in \sqparen{0:M-1}$ and $k \in \sqparen{0:2^n-1}$. We also define $\theta^{\prime}_i \defeq -\pi\paren{\frac{2i}{M}-1}$, $\theta^{\prime\prime}_k \defeq -\paren{k - 2^{n-1}}\frac{2\pi}{2^n}$, and $\theta_{i,k} \defeq \theta^{\prime}_i + \theta^{\prime\prime}_k$ for $i \in \sqparen{0:M-1}$ and $k \in \sqparen{0:2^n-1}$. We first observe that $\e{\jmath \theta_{i, k}} \mathcal{E}_{i, k} = \mathcal{E}_{\frac{M}{2}}$ since multiplication with $\e{\jmath \theta^{\prime}_i}$ rotates the $i$th signal point to $x_{\frac{M}{2}}$ and multiplication with $\e{\jmath \theta^{\prime\prime}_k}$ removes the effect of partition selection for $h$. Secondly, we observe that when $h \in \mathcal{D}_k$, the event $\brparen{x_i \neq \hat{x}\paren{Q\paren{Y},h}}$ is equivalent to $\brparen{Y \notin \mathcal{E}_{i, k}}$  since $\mathcal{E}_{i, k}$ contains all bisectors to which $\sqrt{\snr}h x_i$ is closest for this range of $h$ values. Hence, the following chain of equalities hold:
\begin{eqnarray}
p_{i,k} \hspace{-3mm}&\stackrel{\rm (a)}{=}& \hspace{-3mm}\int_{\mathcal{D}_k}\PR{\sqrt{\snr}h x_i + W \notin \mathcal{E}_{i,k}}f_H\paren{h} dh \nonumber \\
\hspace{-3mm}&\stackrel{\rm (b)}{=}& \hspace{-3mm}\int_{\mathcal{D}_k} \PR{ W \notin \mathcal{E}_{\frac{M}{2}} - \sqrt{\snr}\e{\jmath \theta^{\prime\prime}_{k}}h x_\frac{M}{2}}f_H\paren{h} dh, \nonumber \label{Eqn: pik Derivation 1}
\end{eqnarray}
where (a) follows from the independence of $W$, $H$ and $X$, and (b) follows from above observations and the circular symmetry property of $W$. Let now $z = \e{\jmath \theta^{\prime\prime}_k} h$ above. 
Since multiplication with a unit magnitude complex number is a unitary transformation (i.e., rotation) over the complex plane, we have   
\begin{eqnarray}
p_{i,k} 
\hspace{-2mm}&=& \hspace{-3mm}\int_{\e{\jmath \theta^{\prime\prime}_{k}} \mathcal{D}_k} \hspace{-3mm}\PR{ W \notin \mathcal{E}_{\frac{M}{2}} - \sqrt{\snr}z x_\frac{M}{2}}f_H\paren{\e{-\jmath \theta^{\prime\prime}_{k}} z} dz \nonumber \\ 
\hspace{-2mm}&\stackrel{\rm (a)}{=}&  \hspace{-3mm}\int_{\mathcal{D}_{2^{n-1}}} \PR{ \sqrt{\snr}z x_\frac{M}{2} + W \notin \mathcal{E}_{\frac{M}{2}, 2^{n-1}}}f_H\paren{z} dz \nonumber \\
\hspace{-2mm}&\stackrel{\rm (b)}{=}& \hspace{-3mm}p_{\frac{M}{2}, 2^{n-1}}, \nonumber
\end{eqnarray}
where (b) and (c) follow from the circular symmetry of $H$ \cite{Picinbono94, Koivunen12} and the corresponding definitions for $\mathcal{D}_k$, $\mathcal{E}_{i,k}$ and $p_{i,k}$ for $i \in \sqparen{0:M-1}$ and $k \in \sqparen{0:2^n-1}$. This shows $p\paren{\snr} = 2^n p_{\frac{M}{2}, 2^{n-1}}$. For a circularly-symmetric pdf $f_H\paren{h}$, it is well-known that $rf_H\paren{r\cos \lambda, r\sin \lambda} = \frac{1}{2\pi}f_R\paren{r}$ \cite[Thm. 2.11]{Fang90}.  Hence, switching to polar coordinates, and using the identities $rf_H\paren{r\cos \lambda, r\sin \lambda} = \frac{1}{2\pi}f_R\paren{r}$, $x_\frac{M}{2} = \e{\jmath\frac{\pi}{M}}$ and $\mathcal{E}_{\frac{M}{2}, 2^{n-1}}=\mathcal{E}$, we conclude the proof.

\bibliographystyle{IEEEtran}
\bibliography{icc_ref}
\clearpage
\end{document}